# Spin valve effect by ballistic transport in ferromagnetic metal (MnAs) / semiconductor (GaAs) hybrid heterostructures

Pham Nam Hai[1], Yusuke Sakata[1], Masafumi Yokoyama[1],
Shinobu Ohya[1,2], and Masaaki Tanaka[1,2]

[1]*Department of Electronic Engineering, The University of Tokyo,
7-3-1 Hongo, Bunkyo-ku, Tokyo 113-8656, Japan*

[2]*Japan Science and Technology Agency,
4-1-8 Honcho, Kawaguchi-shi, Saitama 332-0012, Japan*

Abstract

We demonstrate the spin valve effect by ballistic transport in fully epitaxial MnAs ferromagnetic metal / GaAs semiconductor / GaAs:MnAs granular hybrid heterostructures. The GaAs:MnAs material contains ferromagnetic MnAs nanoparticles in a GaAs matrix, and acts as a spin injector and a spin detector. Although the barrier height of the GaAs/MnAs interface was found to be very small, relatively large magnetoresistance was observed. This result shows that by using ballistic transport, we can realize a large spin valve effect without inserting a high tunnel barrier at the ferromagnetic metal / semiconductor interface.

In spintronics applications, metal-based passive devices such as giant magnetoresistance (GMR) head sensors and magnetic random access memory (MRAM) have achieved considerable success. Using the spin degrees of freedom in three-terminal active semiconductor devices is then a natural extension of spintronics research. Recently, novel devices such as spin field-effect transistor (spin FET) [1] and spin metal-oxide-semiconductor field-effect transistor (spin MOSFET) [2] have been proposed as new building blocks for future electronics. Those devices are expected to have the advantages of non-volatility and low power consumption of magnetic devices, as well as high-speed operation of semiconductor devices [3]. The most basic operations of these semiconductor-based spintronic devices require electrical injection of spin-polarized carriers into a semiconductor channel and detection of them by FM electrodes, i.e. the spin valve effect in FM / SC / FM hybrid heterostructure. Unlike the metal-based spin valves, however, this hybrid spin valve structure has a serious problem; due to the large conductivity mismatch between ferromagnetic metals and semiconductors in the diffusive transport regime, the imbalance of injected majority and minority spins in the semiconductor channel becomes extremely small. This "conductivity mismatch" problem has been confirmed in recent theoretical and experimental studies [4-12]. It has been well established that to overcome this problem a tunnel or Schottky barrier has to be inserted at the interface of FM and SC [4-5,13-14]. However, such a high resistance interface is not preferred, because it drastically decreases the current driving capability when used in active transport devices.

In this letter, we show that by using ballistic transport of spin-polarized electrons in a SC channel, we can obtain a large spin valve effect without inserting a high resistance interface. In the case of ballistic transport, the high resistivity of semiconductors is no longer relevant, thus the conductivity mismatch problem may not occur. Consequently, a ballistic semiconductor spintronic device can utilize both the non-volatility of magnetic devices and the high-speed operation of semiconductor devices [15,16]. Indeed, the spin FET and the spin MOSFET are assumed to work under the ballistic transport regime. Nevertheless, realization of ballistic transport in FM / SC / FM spin valve structures has been very challenging. Firstly, the length of the semiconductor channel must be shorter than the mean free path of electrons, that is, it must be a scale of several tens of nanometers or shorter. Secondly, the interface between FM and SC must be very smooth and free of disorders to avoid loss of spin selectivity [15,17], thus any surface treatment techniques such as etching or sputtering should be excluded. Thirdly, because the bias voltage $V_{\text{half}}$, at which the spin valve ratio is reduced by half is typically several hundreds of mV, the Schottky barrier at the FM / SC interface must be low enough to allow ballistic tunneling of electrons from the electrodes to the conduction band of the semiconductor spacer with a small bias voltage.

To satisfy these conditions, we have performed epitaxial growth of FM / SC / FM spin valve structures using molecular beam epitaxy (MBE). Figure 1 shows our spin valve device structure grown on a p+ GaAs(001) substrate. The structure consists of a non-doped GaAs semiconductor layer with the thickness of $t_{GaAs}$ = 10 - 30 nm sandwiched by two ferromagnetic MnAs electrodes. The bottom electrode is a granular thin film, in which ferromagnetic MnAs nanoparticles with size of 5 nm in diameter are embedded in a thin film GaAs matrix (referred to as GaAs:MnAs). The top electrode is a 20 nm–thick type-A MnAs thin film [18]. The growth procedure was as follows. Firstly, we grew a 20 nm-thick Be-doped GaAs buffer layer on a p+GaAs(001) substrate at 580°C. After cooling the substrate temperature to 300°C, we grew a 5 nm–thick $Ga_{0.957}Mn_{0.043}As$ thin film. Then, the structure was annealed at 580°C for 20 minutes in the MBE growth chamber, during which phase separation occurred in the GaMnAs layer and MnAs nanoparticles were formed in the GaAs matrix [19-21]. After that, the substrate temperature was cooled to 300°C again and a GaAs spacer layer with thickness of 10 – 30 nm was grown. Finally, a 20 nm-thick type-A MnAs thin film was grown at 260°C as a top electrode. After completing the growth, post growth annealing was carried out in the growth chamber at 320°C for 10 minutes to improve the structural and magnetic properties of the top MnAs film. By growing the GaAs spacer layer at 300°C, we can suppress diffusion of residual Mn atoms from the GaAs:MnAs electrode to the GaAs spacer while maintaining high crystal quality of GaAs for ballistic transport. The GaAs layers grown at 300°C are semi-insulating and show an electron mobility of 1000 cm$^2$/Vs at room temperature, corresponding to the mean free path of 10 nm. The ballistic transport of electrons through the 300°C-grown GaAs layer at low temperature was confirmed by observing the resonant tunneling effect in AlAs / GaAs / AlAs double barrier diodes with a 300°C-grown GaAs quantum well.

It is worth noting that the overgrowth of high-quality semiconductor layer on top of a FM layer is very difficult. The unique GaAs:MnAs electrode in our spin valve, however, allows the overgrowth of a high-quality and atomically controlled GaAs semiconductor spacer layer [19-21]. Furthermore, the MnAs nanoparticles have been shown to work well as a spin injector and a spin detector [21]. Although the crystal structure of MnAs (NiAs type hexagonal) is different from that of GaAs (zinc-blende), the interfaces between the electrodes and the GaAs spacer are very smooth, as revealed by the streaky patterns of the reflection high energy electron diffraction patterns observed at each layer. The barrier height between MnAs and GaAs in our spin valve structure was found to be very small, as discussed below.

Figures 2 shows the $t_{GaAs}$ dependence of the resistances $R$ of four spin valves with $t_{GaAs}$ = 10, 15, 20, 30 nm plotted in two ways; (a) log($R$)- $t_{GaAs}$ with a bias voltage $V$ of 1 mV, and (b) $R$- $(t_{GaAs})^2$ with $V$ = 50 mV. Assuming direct tunneling of electrons through the GaAs rectangular-type tunnel barrier as shown in the inset of Fig. 2(a), we deduce that the barrier height $\phi$ < 1 meV from the gradient of the log($R$) - $t_{GaAs}$ line by using the WKB approximation. This result is surprising but seems consistent with the fact the Fermi level of MnAs lies above the minimum of the X valley of AlAs semiconductor in MnAs/GaAs/AlAs/GaAs:MnAs magnetic tunnel junctions (MTJs) [21]. The conductance across the MnAs/GaAs interface in our samples is comparable or even better than that of the Co/GaAs interface with $\phi \sim$ 10 meV reported recently [22]. With this very small barrier height, electrons can easily transport from the MnAs electrodes to the conduction band of GaAs by the Fowler-Nordheim (FN) tunneling when the bias is higher ($V$ = 50 mV), as shown in the inset of Fig. 2(b). The FN tunneling current is given by

$$I = \frac{e^3 m_{MnAs} V^2 S}{8\pi m_{GaAs} h \phi t_{GaAs}^2} \exp\left(-\frac{8\pi t_{GaAs}\sqrt{2m_{GaAs}}}{3heV}\phi^{3/2}\right), \qquad (1)$$

where $V$ is the bias voltage, $S$ is the effective area through which the current flows, $h$ is the Plank constant, $e$ is the electronic charge, $m_{MnAs}$ and $m_{GaAs}$ are the effective electron mass of MnAs and GaAs, respectively. When $eV \gg \phi$, the $t_{GaAs}$ dependence of the exponential part of equation (1) is weak, thus $R = V/I$ is proportional to $(t_{GaAs})^2$. Figure 2(b) shows that $R$ measured at 50 mV is proportional to $(t_{GaAs})^2$, revealing the FN tunneling nature of electron transport. In Fig. 2(c), we show the conductance ($G$)–bias voltage ($V$) characteristics of the four spin valve devices. The $G$–$V$ characteristics are clearly different from that of a typical Simmons-type tunnel junction with a rectangular barrier (Ref. [23]), but can be explained by the FN-tunneling based equation (1). In the low bias areas (area I) where the exponential part of equation (1) is smaller than unity, $G$ non-linearly increases with the increasing bias $V$. In the intermediate bias area (area II), $G$ linearly increases with the increasing $V$ since the exponential part of equation (1) has reached unity. Finally, in the high bias area (area III), $G$ increases non-linearly again with the increasing bias. The non-linear increase of $G$ in area III can be attributed to the parallel conduction through the GaAs matrix where MnAs nanoparticles do not exist. This parallel conduction results in the decrease of spin valve ratios at large bias voltages, as discussed later.

Figures 3(a)-(d) show the magnetic-field dependence of the resistance of the spin valve devices measured at 7 K with a bias of 50 mV. The black and red curves are major and minor loops, respectively. The major loops are superposition of the magnetoresistance (MR) components of the ferromagnetic electrodes (gradual change) and the spin valve effect (abrupt jumps of resistance). The hysteresis observed in the minor loops indicates that the resistance jumps correspond to the magnetization reversal of the GaAs:MnAs nanoparticles. If we assume purely diffusive transport of electrons in the GaAs spacer, we can estimate the spin valve ratio by [6]:

$$\text{(Spin Valve ratio)} = 8P^2 \left( \frac{r_{\text{MnAs}}}{r_{\text{GaAs}}} \frac{l^{\text{sf}}_{\text{GaAs}}}{t_{\text{GaAs}}} \right)^2, \qquad (2)$$

where $P$ is the spin polarization of MnAs, $r_{\text{MnAs}}$ and $r_{\text{GaAs}}$ are the products of the resistivity by the spin diffusion length for MnAs and GaAs, respectively, and $l^{\text{sf}}_{\text{GaAs}}$ is the spin-diffusion length of GaAs. Using $P = 0.5$ (Ref. [24]), $r_{\text{MnAs}}/r_{\text{GaAs}} = 10^{-6}$, $l^{\text{sf}}_{\text{GaAs}} = 10$ μm and $t_{\text{GaAs}} = 10$ nm, we get a spin valve ratio of $10^{-6}$. This spin valve ratio expected for purely diffusive transport is 4 orders of magnitude smaller than that of our experimental results. Consequently, the appearance of the spin valve effect of several % in our structures, even when there is no high tunnel or Schottky barriers at the interfaces, indicates that the electron transport is ballistic.

Figure 4(a) shows the bias dependence of the spin valve ratio [defined as $(R_{\text{max}}-R_{H=0})/R_{H=0}$] measured at 7 K, where $R_{\text{max}}$ is the maximum resistance and $R_{H=0}$ is the resistance at zero magnetic field. The sample with $t_{\text{GaAs}} = 10$ nm has the largest spin valve ratio with a maximum of 8.2 % when the bias voltage $V < 120$ mV. However, when $V > 120$ mV, its spin valve ratio decreases rapidly and becomes smaller than that of the sample with $t_{GaAs} = 30$ nm. This is because when $V > 120$ mV, the parallel conduction becomes largest for sample with $t_{GaAs} = 10$ nm but small for the sample with $t_{GaAs} = 30$ nm. Despite the parallel conduction, the spin valve effect was observed up to 300 mV for all samples in our experiments. Figure 4(b) shows the temperature dependence of the spin valve ratios measured with a bias voltage of 50 mV. The spin valve ratios of all samples decreased with increasing temperature, and disappeared at about 90 K. In contrast, the MR component (the gradual change of resistance with the magnetic field sweep) was observed up to room temperature. The absence of the spin valve effect at $T > 90$ K is not due to the temperature dependence of the magnetization of MnAs, since the tunneling magnetoresistance effect of MnAs / GaAs / AlAs / GaAs:MnAs (with $\phi = 5$ nm MnAs nanoparticles) MTJs was observed up to room temperature [25]. The decrease of the spin valve effect with increasing temperature and its disappearance at $T > 90$ K are thus a natural manifestation of the electron transport mechanism in the GaAs channel; when the electron transport changes from ballistic regime to purely diffusive regime with increasing temperature, the conductivity mismatch occurs and destroys the spin valve effect.

In conclusion, we have observed relatively large spin valve effect by ballistic transport in MnAs / GaAs / GaAs:MnAs hybrid heterostructure, even when the barrier height of MnAs/GaAs interface is very small. Our experimental results have shown that by choosing a proper combination of FM / SC materials and using ballistic transport, we can obtain a large spin valve effect without using a high resistance interface. Noting that the silicon technology node has reached 45 nm, it is expected that the ballistic transport of electrons in semiconductors can be utilized to make active semiconductor-based spin devices for future electronics.

This work was supported by JST-SORST/PRESTO, Grant-in-Aids for Scientific Research from MEXT, the Special Coordination Programs for Promoting Science and Technology, and Kurata Memorial Hitachi Sci. & Tech. Foundation. One of the authors (Pham Nam Hai) thanks the JSPS Research Fellowships for Young Scientists.

**Figure legends**

Fig. 1. Schematic structure of our spin valve devices which consist of MnAs thin film (20 nm) / GaAs (10-30 nm) / GaAs:MnAs (5 nm) grown on a p+GaAs (001) substrate.

Fig. 2. Transport characteristics of spin valve structures. (a) GaAs-thickness ($t_{GaAs}$) dependence of the resistance $R$ of four spin valves with $t_{GaAs}$ = 10, 15, 20, 30 nm plotted as log($R$) - $t_{GaAs}$. The inset shows the band diagram of MnAs / GaAs / GaAs:MnAs for the case of direct tunneling of electrons through the rectangular-type GaAs tunnel barrier. The resistances were measured at 7 K with a bias voltage of 1 mV. The black line shows the fitted log($R$) - $t_{GaAs}$ by WKB approximation. The gradient of the fitted line reveals an effective barrier height smaller than 1 meV. (b) $t_{GaAs}$ dependence of the resistance $R$ of the above four spin valves plotted as $R$- $(t_{GaAs})^2$. The inset shows the band diagram of MnAs / GaAs / GaAs:MnAs for the case of FN-tunneling of electrons through the triangular GaAs barrier. The resistances were measured at 7 K with a bias voltage of 50 mV. The dashed line is a guide to the eyes. $R$ is proportional to $(t_{GaAs})^2$, revealing the FN-tunneling nature of electron transport across the MnAs / GaAs interfaces. (c) conductance ($G$) – bias voltage ($V$) characteristics of the four spin valves. The $G – V$ characteristics are clearly different from that of a tunnel junction, but can be explained by equation (1). The $G – V$ characteristics can be divided into three areas: area I where the exponential part of equation (1) is smaller than unity thus $G$ non-linearly increases with the increasing $V$, area II where the exponential part of equation (1) approaches unity thus G is proportional to $V$, and area III where the parallel conduction through the GaAs matrix leads to non-linear increasing of $G$ with $V$. The thin solid line shows the borders of those areas.

Fig. 3. Magnetic-field dependence of the resistance of the spin valve structure with (a) $t_{GaAs}$ = 30 nm. (b) $t_{GaAs}$ = 20 nm. (c) $t_{GaAs}$ = 15 nm. (d) $t_{GaAs}$ = 10 nm. The resistances were measured at 7 K with a bias of 50 mV. The magnetic field was applied in plane along the easy magnetization axis $[\bar{1}\bar{1}20]$ of the MnAs thin film, which is parallel to the GaAs[110] azimuth. The black and red curves are major and minor loops, respectively. The major loops are superposition of the magnetoresistance (MR) components of the ferromagnetic electrodes (gradual change) and the spin valve effect (abrupt jumps of resistance). The hysteresis observed in the minor loops indicates that the resistance jumps correspond to the magnetization reversal of the GaAs:MnAs nanoclusters.

Fig. 4. (a) Bias dependence of the spin valve ratios. The spin valve ratios (defined as $(R_{max}-R_{H=0})/R_{H=0}$) were measured at 7 K. The sample with $t_{GaAs}$ = 10 nm has the largest spin valve ratio with a maximum of 8.2 % when the bias voltage $V$ < 120 mV. When $V$ > 120 mV,

its spin valve ratio decreases rapidly and becomes smaller than that of the sample with $t_{\text{GaAs}}$ = 30 nm due to the parallel conduction. (b) Temperature dependence of the spin valve ratios measured at a bias voltage of 50 mV. The spin valve ratios decreased with increasing temperature, and disappeared at 90 K.

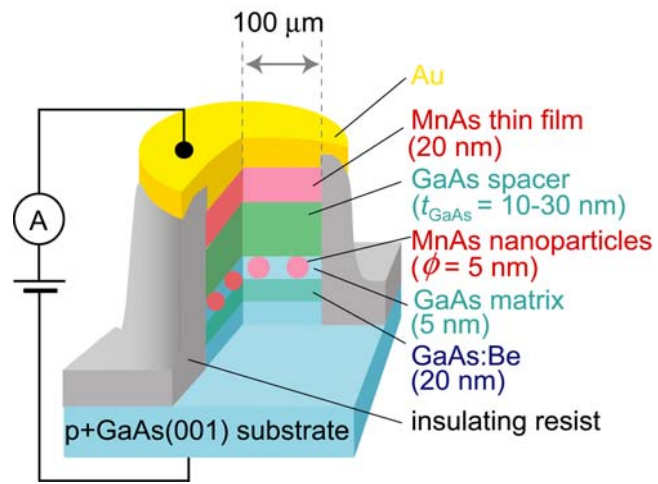

Fig. 1. Hai *et al.*

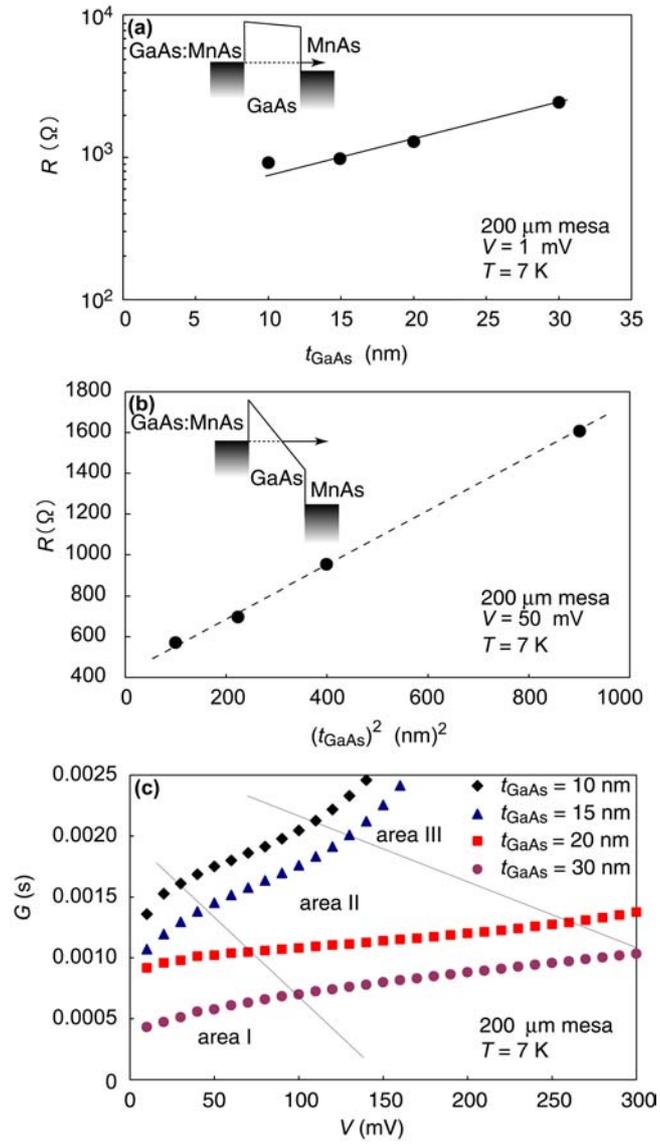

Fig. 2. Hai *et al.*

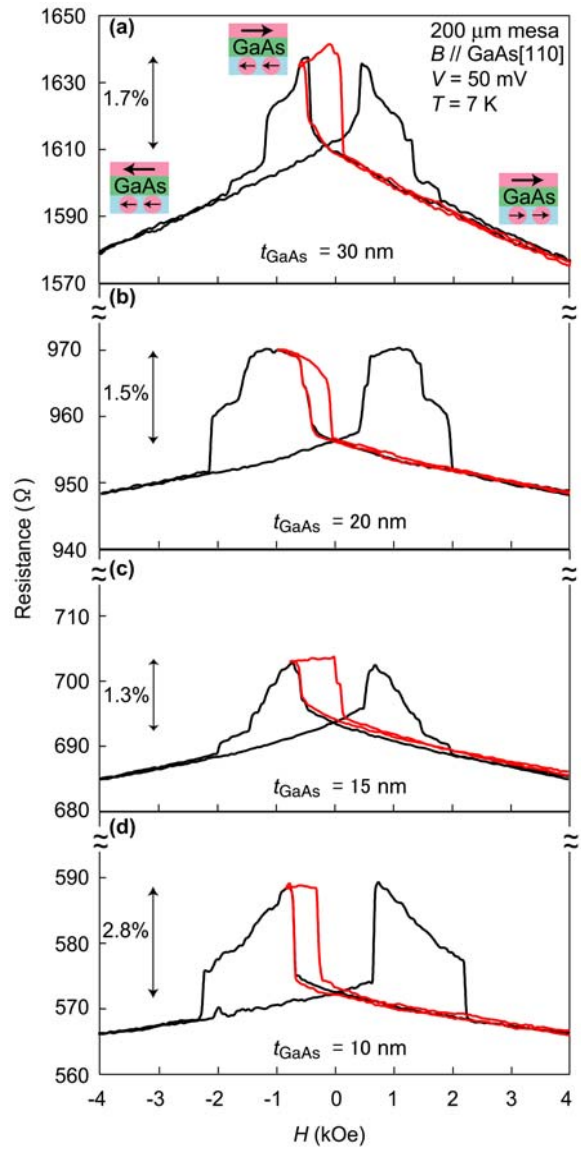

Fig. 3. Hai *et al.*

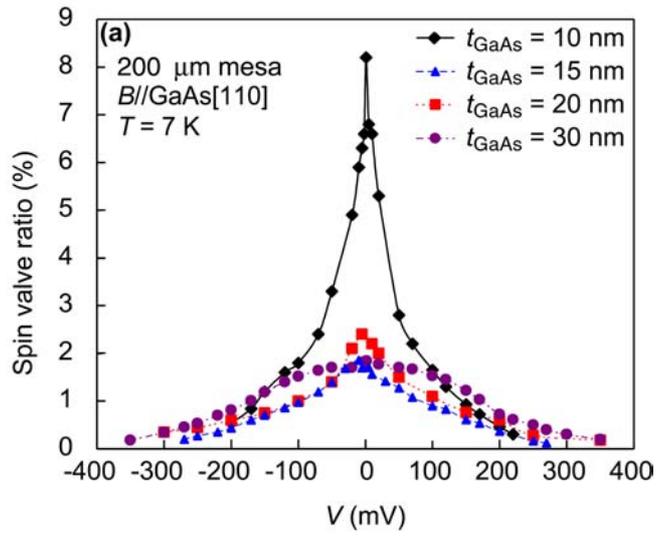
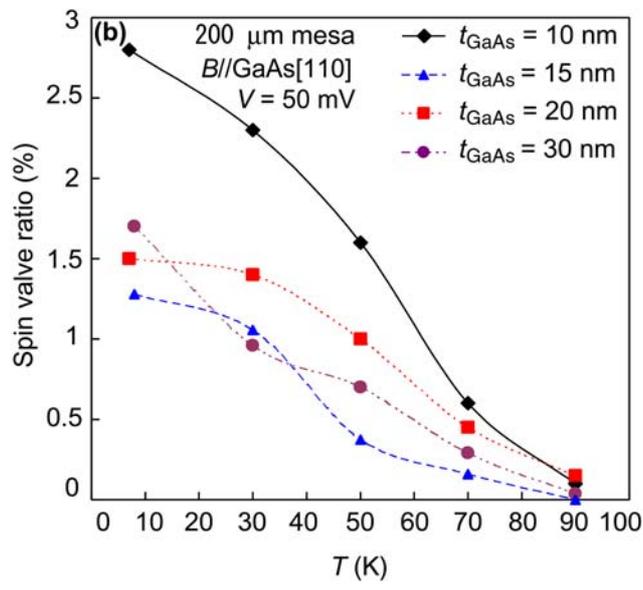

Fig. 4. *Hai* et al.